\documentclass[prb,preprint,showpacs,preprintnumbers,amsmath,amssymb]{revtex4-1}

\usepackage{graphicx}
\usepackage{dcolumn}
\usepackage{bm}
\usepackage{color}

\newcommand{\lunzero}{\ensuremath{\mathrm{L1_{0}}}}
\newcommand{\lunun}{\ensuremath{\mathrm{L1_{1}}}}
\newcommand {\degrees}{\ensuremath{^{\circ}}C}
\newcommand {\degree}{\ensuremath{^{\circ}}}

\newcommand{\alumina}{\ensuremath{\mathrm{Al_2O_3}}}
\newcommand{\qunit}{\AA\ensuremath{\mathrm{^{-1}}}}
\newcommand{\Ta}{\ensuremath{\mathrm{T_a}}}
\newcommand{\Ms}{\ensuremath{\mathrm{M_s}}}

\begin{document}


\title{Origin of perpendicular anisotropy in thin Co/Pt bilayers grown on alumina}
\author{L. Grenet}
\author{C. Arm}
\author{P. Warin} 
\email[corresponding author ]{patrick.warin@cea.fr}
\author{S. Pouget}
\author{A. Marty}
\author{A. Brenac}
\author{L. Notin}
\author{P. Bayle-Guillemaud}
\author{M. Jamet}

\affiliation{CEA, INAC, SP2M, NM, UMR-E 9002, 17 rue des martyrs, F-38054 Grenoble, France.}
\affiliation{Universit\'e Joseph Fourier, SP2M, UMR-E 9002 F-38041 Grenoble, France.}
\date{\today}

\begin{abstract}

We investigate in this paper the origin of perpendicular anisotropy in Co (1.6 nm)/Pt (3.0 nm) bilayers grown on alumina and annealed up to 650$^{\circ}$C. Above 350$^{\circ}$C, all layers exhibit perpendicular anisotropy. Then coercive fields increase linearly with annealing temperature following two different rates: 0.05 T/100$^{\circ}$C below 550$^{\circ}$C and 0.8 T/100$^{\circ}$C above. By making careful structural characterizations using x-ray diffraction and transmission electron microscopy, we demonstrate the presence of short range correlation of L1$_{1}$ type below 550$^{\circ}$C whereas above 550$^{\circ}$C, L1$_{0}$ chemical ordering is observed. We conclude that perpendicular anisotropy observed in Co/Pt bilayers grown on alumina and annealed may not only be due to interface anisotropy as usually invoked but also to CoPt alloying and chemical ordering that take place during post-growth annealing.

\end{abstract}
\maketitle

\section{Introduction}

Materials displaying a large Perpendicular Magnetic Anisotropy (PMA) are very attractive for the media storage industry. Indeed their very strong magnetic anisotropy may allow to build smaller storage elements while still overcoming the superparamagnetic limit. Hence they allow a larger areal density\cite{Weller2000}. PMA is although very appealing for Magnetic random access memory (MRAM) industry for its possibility to reduce the bit size. Spin transfer torque (STT) is also easier to implement because current density required to reverse the storage layer is considerably smaller for materials exhibiting PMA\cite{Nakayama08}.

PMA is found in many magnetic materials: Co-based multilayers\cite{chappert1986}, chemically ordered (\lunzero) Fe or Co based alloys (CoPt, FePt, CoNi). Methods have also been designed to create PMA in materials with in-plane anisotropy. Chappert and collaborators used irradiation by He+ ions at 30 keV in Pt/Co/Pt \cite{Chappert1998} to induce PMA. Monso and collaborators \cite{Monso2002} demonstrated that on very thin layers of Co sandwiched between Pt and Al, oxidation of the aluminum layer leads to a reorientation transition of the magnetization of the Co layer from in plane to out-of-plane. Manchon \textit{et al.} demonstrated that the appearance of PMA in Pt/Co/\alumina is triggered by an optimum oxidation of the aluminum layer. When it is underoxidized the magnetization is in-plane. When the aluminum layer is overoxidized, CoO starts to form. The cobalt layer is no longer 100 \% remanent\cite{Manchon2008}.

However, all the methods presented above are difficult to implement in magnetic tunnel junction (MTJ) where PMA is most desired for reasons mentioned above. Indeed \lunzero{} FePt can be grown on tunnel barrier of MgO but crystalline quality is not as good as FePt layer grown on a Pt buffer\cite{Person2007}. Co based multilayers, or Co ultrathin layers sandwiched between layers of noble metals require a thick noble metal buffer that is detrimental for many spintronic application. $\mathrm{He^+}$ irradiation can cause damages in the tunnel barrier. Monso \textit{et al.} solution works only for the bottom layer of the MTJ. Therefore a method compatible with tunnel barrier engineering is needed.

Recently, Nistor and collaborators \cite{Nistor2009} have shown that in $\mathrm{SiO_2}$/Co(15\AA)/Pt(30\AA) or \alumina/Co(15\AA)/Pt(30\AA), it is possible to obtain a high PMA of the top electrode after annealing at 400 \degrees. In this work, PMA was atributed to interface anisotropy as a consequence of the strong Co-O hybridization at the cobalt/alumina interface. These results further triggered a large interest. We have used the same method to grow perpendicularly magnetized electrodes (\alumina/Co(15\AA)/Pt(30\AA)) on top of an optical spin detector (a spin-led\cite{Fiederling99}) to inject and detect spin polarized electron into silicon at remanence \cite{Grenet2009}. 

Major advances have been made on this topic by groups who wanted to engineer layers displaying PMA using MgO/CoFeB layers \cite{Ikeda2010,Jung2010} that are under the spotlight because they allow for large tunnel magnetoresistance. However the origin of PMA in these oxide-metals systems is still unclear.

This article deals with another explanation to the origin of PMA in the \alumina/Co/Pt system. We start by studying the evolution of the magnetic properties of the \alumina/Co/Pt as a function of annealing temperature. X-ray diffraction and Electron-energy loss spectroscopy are then used to study the consequences of annealing  on the structural properties. The first section details the experimental techniques used for the study. The second part describes the experimental results that are discussed in the third part.



\section{Experimental techniques}

\alumina/Co/Pt is grown on Si (100). After removing SiO$_2$ native oxide with hydrofluoric acid and rinsed in de-ionized water for 15 minutes, a 3.2 nm thick \alumina~ layer is grown at room temperature by using radio frequency magnetron sputtering. The evaporation rate of the alumina is about 0.25 \AA.s$^{-1}$. Samples are then moved to another deposition chamber with a short-time exposure to outside atmosphere. The Co(1.5 nm)/Pt(3 nm) ferromagnetic layer is grown at room temperature using a direct current sputtering magnetron. For both metals, the deposition rate is about 1 \AA.s$^{-1}$. After deposition, the whole structure is annealed in a separated oven under a secondary vacuum of 10$^{-5}$ mbar. The procedure consists in 1 min. annealing at 115$^\circ$ C followed by 90 min. annealing at the temperature set point $\mathrm{T_a}$. To avoid spurious effects due to time variation of the sputtering growth rate, all the samples presented in this paper have been grown simultaneously.

Three experimental techniques have been used for magnetic measurements: polar magneto-optical Kerr effect (MOKE), extraordinary Hall effect (EHE) and superconducting quantum interference device (SQUID). The SQUID setup further permits both in-plane, out-of-plane and temperature dependent absolute magnetization measurements.

X-ray diffraction and reflectivity were used to characterize the evolution of Co and Pt layers upon increasing annealing temperature.
A conventional X'Pert Panalytical two-circle x-ray diffractometer with Co $\mathrm{K_{\alpha}}$  radiation was used to perform $\theta-2\theta$ and $\omega$ scans. The incident beam divergence slit was set at 0.25\degree, while the diffracted beam divergence was limited by a 0.27\degree{} long plate collimator. 
A Seifert diffractometer with Cu $\mathrm{K_{\alpha}}$  radiation equipped with an eulerian cradle was used for the pole figures measurements. 


\section{Evolution of magnetic properties with annealing temperature}

\begin{figure}[hbt!!]
\begin{center}
\includegraphics*[angle=0,width=0.5\textwidth]{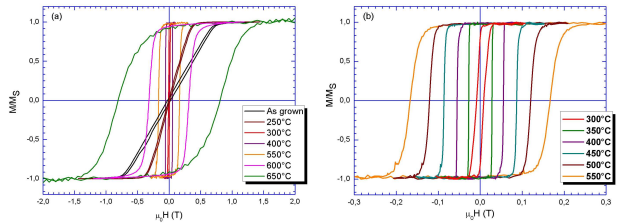}
\caption{(a) Hysteresis loop of the $Al_20_3/Co/Pt$ layer for different annealing temperature measured with the field applied perpendicular to the plane and at room temperature (b) Zoom at small magnetic field.\label{Hysteresis}}
\end{center}
\end{figure}

Fig.\ref{Hysteresis} shows several hysteresis loops with the applied field perpendicular to the film of \alumina/Co/Pt as a function of the annealing temperature. As-grown sample exhibits in-plane magnetization, which is expected from Co very thin film. Therefore the hysteresis loop is typical of hard axis behavior. For an annealing temperature of 250 \degrees{} the same behavior is observed. For an annealing temperature of 300 \degrees, it seems that the normal to the plane of the film becomes the easy axis but the anisotropy is not strong enough to overcome the demagnetising field. Starting from an annealing temperature of 350 \degrees{} and until 550 \degrees{} all samples exhibit 100 \% remanence. This is a clear signature of a strong perpendicular anisotropy. 

\begin{figure}[hbt!!]
\begin{center}
\begin{tabular}{cc}
\mbox{\mbox{\includegraphics*[angle=0,width=0.25\textwidth]{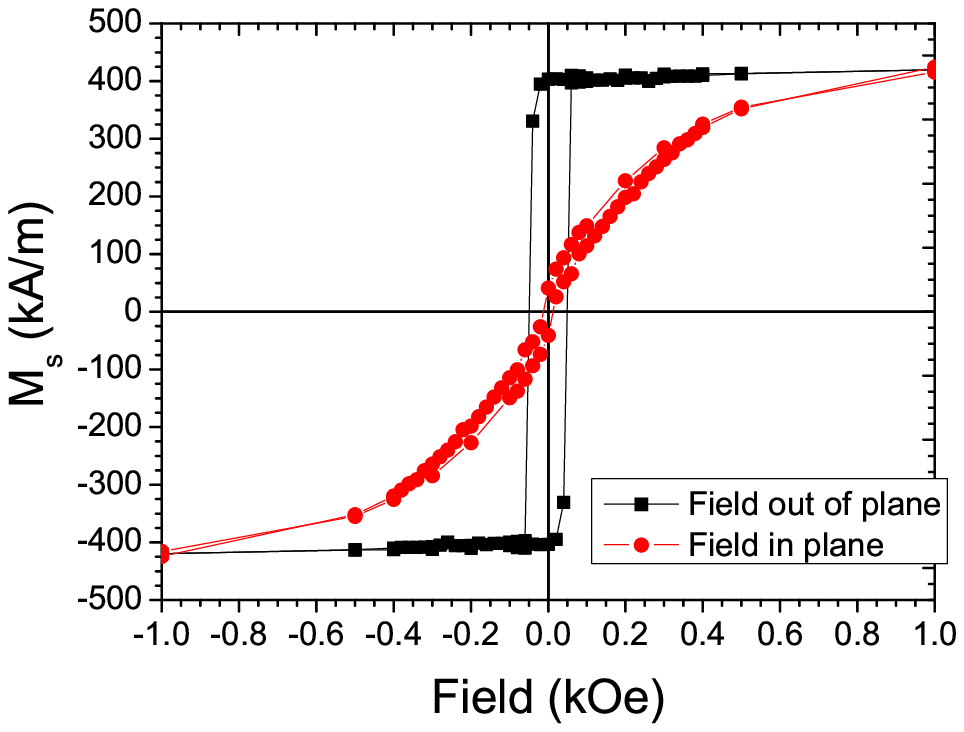}}\includegraphics*[angle=0,width=0.25\textwidth]{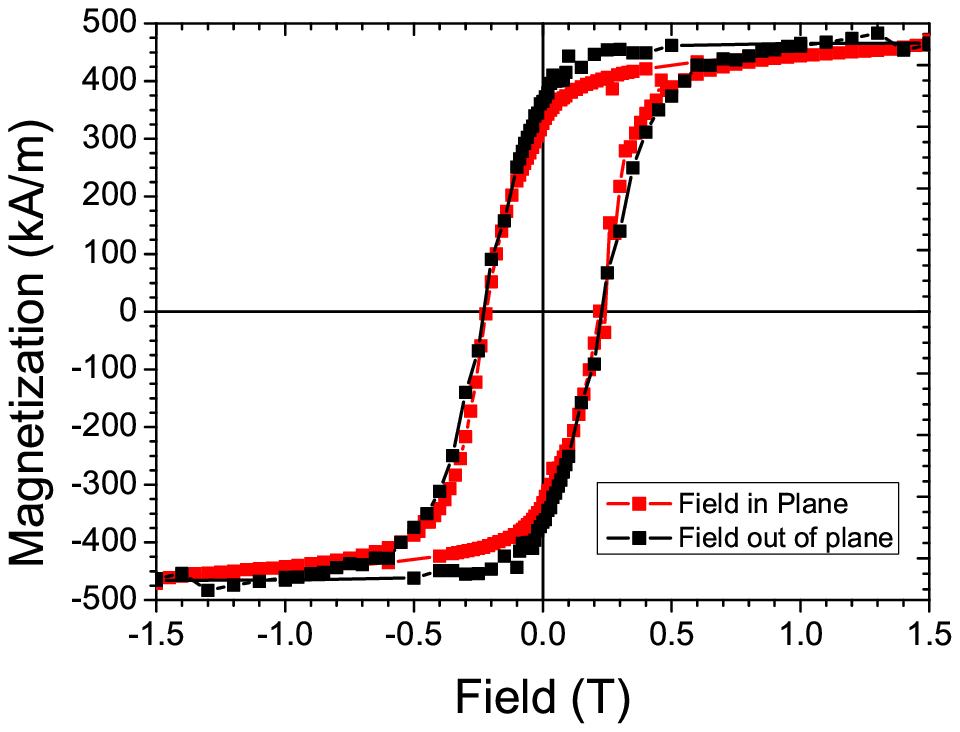}}&
\end{tabular}
\caption{a)Hysteresis loop for the sample annealed at 400 \degrees{} with the field applied in plane and out of plane. b) Hysteresis loop for \Ta=600\degrees{}. Note that in plane and out of plane loops look very similar.\label{Hyst}}
\end{center}
\end{figure}

Fig.\ref{Hyst}a shows the hysteresis loop recorded at 300 K for sample annealed at 400 \degrees{} using SQUID magnetometry. The in-plane loop presents a hard axis behavior. We have calculated the uniaxial anisotropy by calculating the surface between the out of plane loop and the in plane loop. We obtained $\mathrm{K_u}=1.4\ 10^6\ \mathrm{J.m^{-3}}$. This is much larger than the anisotropy of pure Co, but of the order of magnitude of Pt/Co/Al layers of Monso \textit{et al.}.\cite{Monso2002}

Fig.\ref{Hyst}b shows the hysteresis loop recorded at 300 K for sample annealed at 600 \degrees{} with the field applied in plane (SQUID) and out-of-plane. It clearly demonstrates that the easy axis of magnetization has rotated away from the normal to the film plane. Coercive field is around 0.2 T for in-plane and out-of plane applied field. SQUID measurement is an absolute measurement of the magnetization contrary to Kerr effect or AHE. After subtracting the diamagnetic component of the substrate it is possible to extract the saturation magnetization. We obtained \Ms=1150 kA/m for \Ta=400 \degrees{} and \Ms = 1427 kA/m for \Ta= 600 \degrees{} for 1.6 nm of Co. \Ms\ for \Ta=400\degrees{} is much less than bulk Co magnetization at saturation. However this may be due to the very small thickness of the layer in our case.



%
%

\begin{figure}[hbt]
\begin{center}
\includegraphics*[angle=0,width=0.5\textwidth]{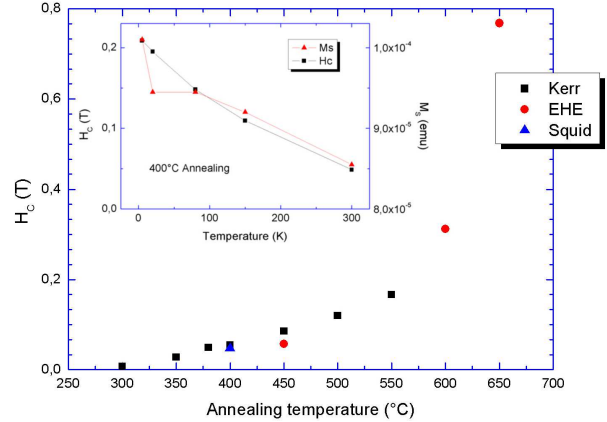}
\caption{(b) Coercive field variation of a Al$_2$O$_3$/Co/Pt layer as a function of annealing temperature. Coercive field are extracted from MOKE, EHE and SQUID measurements. Inset shows the temperature dependence of the coercive field for a 400 \degrees{} annealed sample.\label{Aimantation}}
\end{center}
\end{figure}

In Fig.\ref{Aimantation} the coercive field is plotted as a function of annealing temperature $\mathrm{T_a}$. From 250 \degrees{} the coercive field slowly increases from 0 T to reach 0.18 T at 550 \degrees. Above that temperature there is a dramatic increase of the coercive field that reaches nearly 0.8 T for the sample annealed at 650 \degrees. This rapid increase of the coercive field, coupled with the hysteresis loop at \Ta=600\degrees{} motivated us to perform X-ray diffraction and transmission electron microscopy on these layers to elucidate the consequences of annealing on the structural properties of these layers.

\section{Structural analysis}

\subsection{X-ray diffraction}

Diffraction is used to probe qualitatively and quantitatively several features such as  cristallinity, structure, texture, lateral and vertical crystallite sizes and mosaicity. Considering the $\stackrel{\rightarrow}{\mathrm{k_i}}$ and $\stackrel{\rightarrow}{\mathrm{k_f}}$ wave vectors respectively characterizing the incoming and outcoming beams, it is essential to keep in mind that the direction probed during the measurement is the direction of the momentum transfer vector $\mathrm{\stackrel{\rightarrow}{Q} = \stackrel{\rightarrow}{k_i} - \stackrel{\rightarrow}{k_f}}$ , and the probed length scale is of the order of the inverse of its modulus. It then comes out that a symmetrical $\theta-2\theta$ measurement (for which the incident angle is half of the diffusion one) gives access to the crystallographic direction and the characteristic correlation size  perpendicular to the layers, $\mathrm{L_{\bot}}$. Rocking curves, in which the incident angle only is varied thus transferring a momentum component in the plane of the layer, probe the mosaicity of the atomic planes perpendicular to the growing direction together with the $\mathrm{L_{\|}}$ in-plane correlation length.

\subsubsection{$\theta-2\theta$ measurements}

\begin{figure}[hbt]
\begin{center}
\includegraphics*[angle=0,width=0.5\textwidth]{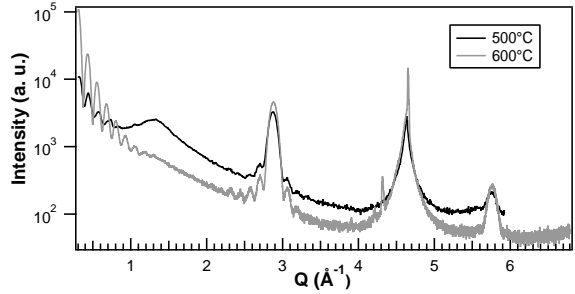}
\caption{$\theta/2\theta$ diffractograms measured with samples annealed at 500 \degrees{} and 600 \degrees. For sake of clarity, the 500 \degrees{} curve is slightly offset .\label{Iq500600}}
\end{center}
\end{figure}

Fig.\ref{Iq500600} shows $\theta-2\theta$ spectra obtained with two samples respectively annealed at 500 \degrees{} and 600 \degrees{} for $2\theta$ values between 5\degree{} and 150\degree, corresponding to momentum transfer values between 0.31 \qunit{} and 6.78 \qunit. In the small angle region corresponding to x-ray reflectivity, the oscillations period  $\Delta\mathrm{Q}=$ yields a thickness $e=2\pi /\Delta Q = 51,5 \pm 0,1$ \AA. The persistence of oscillations up to a diffusion angle of 20\degree{}(or 1.20 \qunit) traduces high quality interfaces with low roughness. A broad peak can be distinguished around Q=$\|\stackrel{\rightarrow}{Q}\|$ = 1,33 \qunit. Further on, at Q = 2,88 \qunit{} and Q = 5,76 \qunit{} two peaks are present which could correspond to the (111) and (222) reflections of either the CoPt  disordered A1 phase or the CoPt ordered \lunzero{} phase. The (111) peak displays \textit{sinc} function oscillations due to the presence of abrupt interfaces. The peak centered at Q = 4,66 \qunit{} corresponds to the (400) reflection of the Si substrate.

\begin{figure}[hbt!!]
\begin{center}
\begin{tabular}{cc}
\mbox{\includegraphics*[angle=0,width=0.5\textwidth]{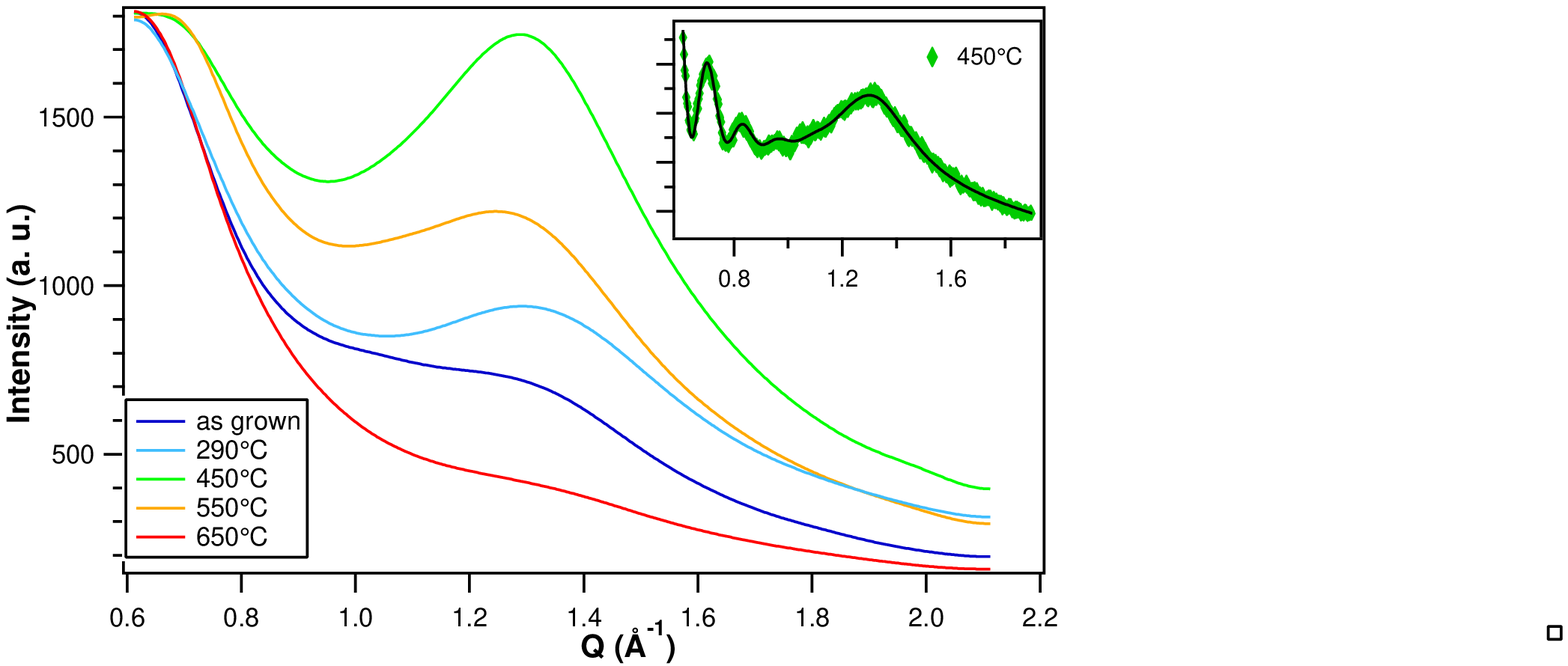}}&
\end{tabular}
\caption{Annealing temperature dependence of the low-Q part of the  $\theta/2\theta$  diffractograms. The curves were smoothed and normalized. Inset: Raw data measured with the 450 \degrees{} annealed sample, fitted with the sum of a dumped sinc square function and a lorentzian.\label{lowq003}}
\end{center}
\end{figure}

Fig.\ref{lowq003} shows the temperature dependence of the low-Q part of $\theta-2\theta$ spectra. The spectra were smoothed and normalized to a common level at $2\theta$ = 5\degree. The main feature is the non monotonous variation of the broad peak intensity. Starting from a non-zero value for the as-grown sample, it increases until reaching its maximum for 450 \degrees{} and decreases for higher annealing temperatures. It is no more visible for 650 \degrees. 
In the insert of Fig.\ref{lowq003}, the experimental data of the 450 \degrees{} annealed sample were fitted using the sum of a dumped squared $sinc$ function and a lorentzian function. It yields a peak position at $1.31 \pm 0.01$ \qunit{} and a halfwidth at half maximum equal to $0.18 \pm 0.01$ \qunit{} corresponding to a correlation length of the order of 35 \AA. This broad peak could be attributed to the presence of \lunun{} type CoPt crystalline order with the (111) direction of the rhombohedron perpendicular to the layers, as already observed in Co-Pt films \cite{Iwata1997}. This type of arrangement can be described as (111) stacking of distorted cubes presenting a chemical order along the diagonal. The \lunun{} crystallographic phase is often characterized by its hexagonal lattice, the (111) reflection thus becoming (003). The (006) reflection is expected to be close to the A1(111) peak. From the position of the fitting lorentzian peak we can deduce a value for the $c$ lattice parameter of the hexagonal cell: $c$ = 14.39 $\pm$ 0.01 \AA. 

\begin{figure}[hbt!!]
\begin{center}
\begin{tabular}{cc}
\mbox{\includegraphics*[angle=0,width=0.5\textwidth]{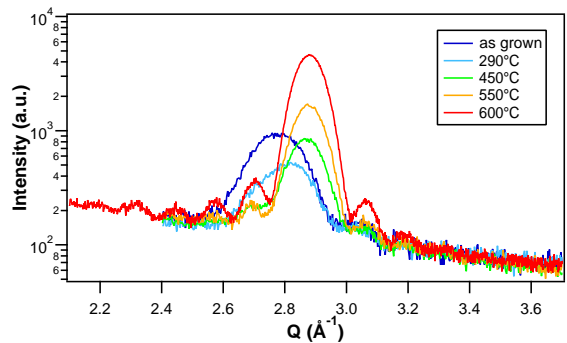}}&
\end{tabular}
\caption{Close-up around the (111) peak from the $\theta/2\theta$ curves for different \Ta.\label{l11006}}
\end{center}
\end{figure}

The evolution of the 2.88 \qunit{} peak (see Fig.\ref{Iq500600}) on the annealing temperature is presented in Fig.\ref{l11006}. Upon decreasing the annealing temperature, the $sinc$ function oscillations disappear and the peak shifts towards lower Q values. 

\begin{figure}[hbt!!]
\begin{center}
\begin{tabular}{cc}
\mbox{\includegraphics*[angle=0,width=0.5\textwidth]{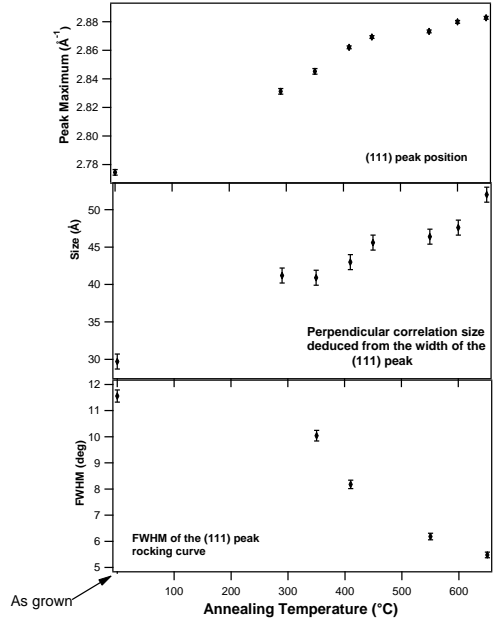}}&
\end{tabular}
\caption{Evolution as a function of \Ta{} of the (111) peak position, the crystallite size perpendicular to the layer and the rocking curve width. \label{peakP2}}
\end{center}
\end{figure}

The 650 \degrees, 600 \degrees{} and 550 \degrees{} data were fitted with a squared \textit{sinc} function.
For lower annealing temperatures the data were fitted with the sum of two Gaussian functions. 
Fig.\ref{peakP2}a and b shows the temperature evolution of the main peak position and of the crystallite size deduced from its width. This size varies from e = 52 $\pm$ 2 \AA{} for the 650 \degrees{} annealed sample, which corresponds to the sum of the expected Co and Pt layer thicknesses, to e = 31 $\pm$ 2 \AA{} for the as-grown sample, corresponding to the expected thickness of the Pt layer. For this sample, the peak position is the one expected for the (111) reflection of the Platinum cubic phase. It increases upon increasing the annealing temperature and reaches the position expected from the Vegard law for a three to two Pt\textbackslash Co proportion of the two elements.

\subsubsection{Rocking curves}

Rocking curves of the main peak were recorded for different annealing temperatures \Ta. The experimental curves were fitted with the sum of a pseudo-voigt function and a linear background. The results of these fits show an increase of the value of the $\eta$ parameter of the pseudo-voigt function (which characterizes the weight of lorentzian with respect to Gaussian) for lower annealing temperatures. 
We checked that the values of the full width at half maximum (FWHM) are not much influenced by the value of $\eta$; figure \ref{peakP2} displays the evolution of this FWHM with the annealing temperature. As already mentioned, the width of a rocking curve is due to both mosaicity and lateral correlation length, which do not have the same $2\theta$ dependency. Measuring rocking curves for different orders of the symmetrical reflection allows discriminating between both effects (Williamson Hall plot \cite{Metzger1998}). The quantity  $\beta_{\omega}\sin\theta/\lambda$ (where $\beta_{\omega}$,$\theta$  and $\lambda$  respectively correspond to the integral breadth, the Bragg angle and the wavelength ) varies as a linear function of $\sin\theta/\lambda$; whereas the slope directly gives the mosaicity, the ordinate at origin is inversely proportional to the lateral correlation length $\mathrm{L_{\|}}$.  In the case of the 600$^{\circ}$C annealed sample, the (111) and (222) rocking curve fits were reliable enough to allow such an analysis; but it has to be noticed that the expected precisions for the determined values are poor, due to the fact that we dispose of only two (hhh) reflections. We obtain a lateral correlation length value $\mathrm{L_{\|}}$= 83 $\pm$ 20 \AA{} and a mosaicity value of 5.7 $\pm$ 0.5\degree.

If we summarize at this point, it appears that short range correlations of \lunun{} order type develop and coexist with the A1 or \lunzero{} lattice. These correlations are maximum at an annealing temperature of the order of 450 \degrees. 

\begin{figure}[hbt!!]
\begin{center}
\begin{tabular}{cc}
\mbox{\includegraphics*[angle=0,width=0.5\textwidth]{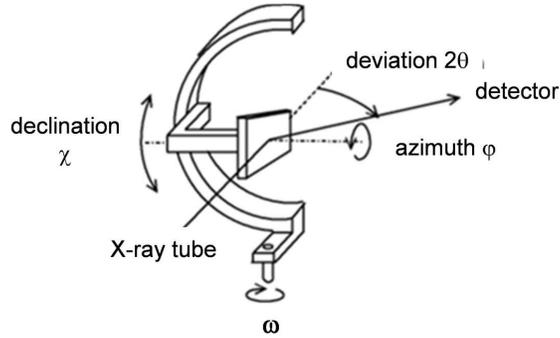}}&
\end{tabular}
\caption{Angle definitions for the pole figure measurement.\label{XRDangle}}
\end{center}
\end{figure}

\subsubsection{Pole figures}

\begin{figure}[hbt!!]
\begin{center}
\begin{tabular}{cc}
\mbox{\includegraphics*[angle=0,width=0.5\textwidth]{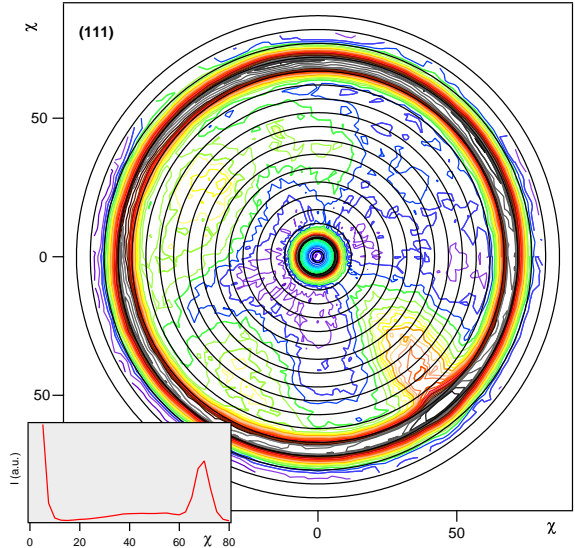}}&
\end{tabular}
\caption{[111] pole figure. The ring close to $\chi$=70.5\degree{} proves the polycristalline nature of the sample in the layer plane. In insert, radial integration of intensity as a function of declination $\chi$. It shows a well defined peak close to $\chi$=70.5\degree{} which corresponds to the angle between two [111] equivalent directions. \label{pole111}}
\end{center}
\end{figure}

$\theta-2\theta$ measurements on such a textured sample cannot allow to discriminate between A1 and \lunzero{} phases. Pole figure measurements were performed on the 650 \degrees{} annealed sample. They consist in fixing $2\theta$ and $\omega$, and varying the azimuth and declination angles respectively known as $\phi$ and $\chi$ (see Fig.\ref{XRDangle}).
The (111) pole figure is shown in Fig.\ref{pole111}. The concentric black circles correspond to different values of the declination angle $\chi$ varying from 0 to 90\degree{} by step of 5\degree. Different positions on a given circle correspond to different values of the azimuth angle $\phi$. $2\theta$ was set at the expected value of the Bragg angle for the (111) reflection, and $\omega=\theta$.  

\begin{figure}[hbt!!]
\begin{center}
\begin{tabular}{cc}
\mbox{\includegraphics*[angle=0,width=0.5\textwidth]{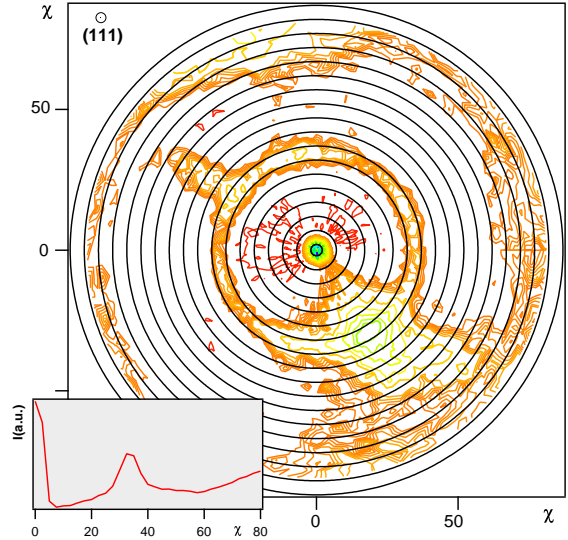}}&
\end{tabular}
\caption{[110] pole figure. A ring is present close to a declination value of 35\degree, which corresponds to the angle between [111] and [110] directions. This proves the presence of \lunzero{} phase.\label{pole110}}
\end{center}
\end{figure}

In the case of a cubic single crystal layer, the pole figure is expected to display the characteristic three fold symmetry of the <111> direction, i.e. three spots corresponding to the three (-111), (1-11) and (11-1) reflections. This is not what is observed in Fig 7. Instead of three different spots, a ring is present for a declination value which is close to the angle between the layer normal [111] and any of the three directions [-111], [1-11] and [11-1]. This result proves that the CoPt crystallites, which are [111] textured, are randomly oriented in the layer plane.
In order to discriminate between A1 and \lunzero{} phases the \lunzero{} [001] and [110] pole figures were measured. In the case of the A1 phase, no signal from the CoPt layer is expected as these two reflections do not exist due to lattice symmetry. Fig.\ref{pole110} shows the [110] pole figure, which evidences a diffraction ring at the expected declination values for the \lunzero{} phase (angle between [110] and [111] direction). The inserts display the $\chi$ dependence of the scattered intensity summed over the azimuth angle (radial integration), excluding the angular domains corresponding to spurious peaks. This result proves the presence of \lunzero{} CoPt phase in the 650 \degrees{} annealed sample.

\subsection{Transmission electron microscopy}

%

\begin{figure}[hbt!]
\begin{center}
\begin{tabular}{cc}
\mbox{\includegraphics*[angle=0,width=0.4\textwidth]{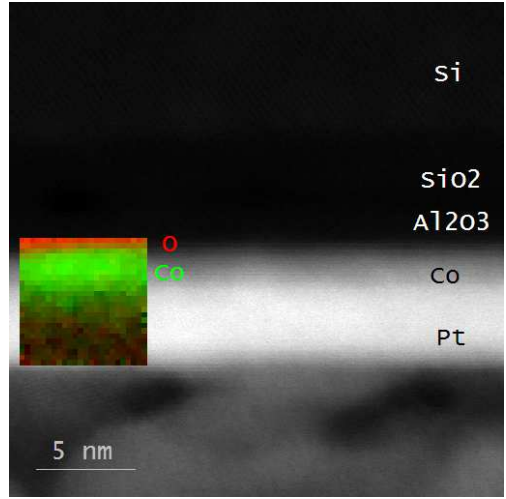}}&
\end{tabular}
\caption{STEM/HAADF picture of the Si/\alumina/Co/Pt layer. In insert chemical map of Co (red)and O(green) in this area.\label{haadf}}
\end{center}
\end{figure}

Electron Energy loss spectroscopy in STEM (scanning transmission electron microscopy) mode has been performed on these samples using a FEI-titan cs-probe corrected working at 300kV. Co-$\mathrm{L_{2,3}}$, O-K, Pt-$\mathrm{M_{4,5}}$, and Si-K edges have been recorded across the interfaces in order to extract chemical profiles. Figure \ref{haadf} shows for the as-grown sample, a STEM/HAADF(High angle annular dark field) image with superimposed a color map showing in green the O and in red the Co distribution in the \alumina/Co/Pt area. The lighter contrast on the STEM image is localized on the Co/Pt layers :intensity is proportional to the atomic number leading to a very light contrast on the Pt layer. From the chemical maps, it is clear that there is no Cobalt oxide and that the Co layer is localized on top of the \alumina. 

\begin{figure}[hbt!]
\begin{center}
\begin{tabular}{cc}
\mbox{\includegraphics*[angle=0,width=0.4\textwidth]{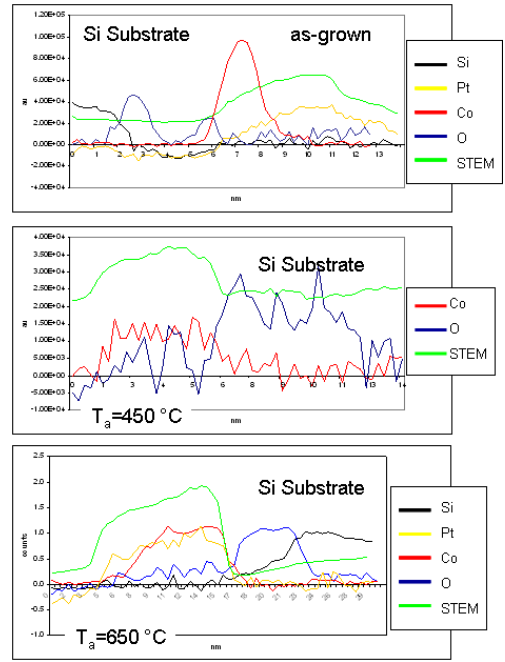}}&
\end{tabular}
\caption{EELS profiles for several chemical elements recorded at three different annealing temperatures. top) as-grown sample. middle) Sample with \Ta=450\degrees. Note that the Si substarte is on the right. Note also that the two metallic layers are intermixed. bottom) Sample annealed at \Ta=650\degrees. \label{eels}}
\end{center}
\end{figure}

Figure \ref{eels} shows a series of chemical profiles acquired across the interfaces of the as-grown, 450\degrees{} and 650\degrees{} annealed samples. The green profile is the STEM/HAADF profile and allows defining the width of the Co/Pt layer. It is clear that annealing increases the width of the Co signal due to diffusion of Co in Pt. At 450\degrees{} and 650\degrees{}, Co is found on the whole thickness superimposed on STEM signals. For 650\degrees{} Co and Pt profiles are in good accordance showing full intermixing. Moreover, no O signal is detected in this layer (above noise level). These results are in good agreement with the XRD measurements. 

\section{Discussion}

Let us first summarize all the information gathered concerning the consequences of annealing on \alumina/Co/Pt thin films. We have observed that the coercive field increases slowly with \Ta{} until \Ta{} reaches 500 \degrees{}. At that point the slope changes abruptly. We also observe that the magnetization at saturation for \Ta=600 \degrees{} is not compatible with pure Co.
Moreover in-plane hysteresis loop recorded on this sample show that the hard axis is no more along the film normal.

XRD shows the presence of the peak labeled \emph{111} for cubic symmetry normally associated with chemical ordering. Let us remind that metal growth on oxide is in most cases along the direction perpendicular to the densest plane\cite{Poate78}. In the case of Co it could either be (0001) hcp or (111) fcc. In ultrathin films, it is the fcc symmetry that is the most observed. The presence of this (111) peak is the signature of CoPt alloy, chemically ordered in the \lunun{} phase. This chemically ordered phase is an explanation to the strong PMA observed on our film. It is also an alternative explanation to the one given by Nistor \textit{et al.} \cite{Nistor2009}. \textit{In their work, Nistor \textit{et al.} attributed the existence of PMA to interfacial magnetic anisotropy induced by Co-O bonds. The density of these bonds is increased by an optimal annealing. \textit{Ab initio} calculations also predict the appearance of PMA at the interface of Fe/MgO\cite{Yang2010-condmat}.}

Above 450 \degrees{} the intensity of (111) peak associated with the \lunun{} phase decreases. At \Ta=600 \degrees{} it has completely disappeared. The hysteresis loop recorded for this sample has an altogether different aspect than the one recorded at 400 \degrees. It shows that the (111) axis is no more the easy axis. Pole figure recorded at Ta=650 \degrees{} shows that the sample exhibits now a \lunzero{} phase. Considering that our samples are polycrystalline (see Fig.\ref{pole111}) with a (111) texture, the transformation from a \lunun{} phase  with 111 easy axis to a \lunzero{} phase with grains having randomly oriented 001 easy axis around the 111 axis explains the shape of the hysteresis loop for \Ta=600 \degrees{}(Fig\ref{Hyst}b). Hysteresis loop with in plane and out-of plane applied field are very similar because no one is recorded either along an easy axis or a hard axis. This explains the strong increase in the coercive field observed above 550 \degrees.

We then discuss the mechanism that induces the \lunun{} ordering. The growth of Co on an oxide surface presents usually a (111) texture. According to Maurice et al \cite{Maurice1999}, full covering is not reached before an equivalent thickness of 3.2 nm of Co is deposited. The deposition rate in their study is 0.1 $\mathrm{\mathring{A} . s^{-1}}$, that is ten times smaller than our. However, in the case of the growth on a oxide surface, the dominant process is the growth on the defect of the surface. Therefore this difference in the deposition rate should not alter considerably the conclusion of Maurice et al. In their study, for a thickness of Co of 1.5 nm, it is clear that Co clusters have coalesced and the percolation threshold is close to that thickness. Platinum deposition on this layer certainly fills the gaps in the Co layer, because the roughness measured on the as-grown film is 4.2 \AA (measured on a 5 $\mu$m by 5 $\mu$m surface). The texture of the platinum deposited is also (111), because Pt atoms (or adatoms) either are on an oxide surface adopting a (111) texture or are on Co atoms which have the same texture. The surface tension of Pt atoms sitting on the coalesced clusters is higher because of their curvature. We may expect interdiffusion of Co and Pt in order to decrease the surface tension. This might explain the small bump at Q = 1,33 \qunit{} on the 111 peak observed on the as grown film, that is a trace of the \lunun{} ordering.

The effect of annealing is to increase the Co-Pt interdiffusion due to kinetics effects. That has been reported in several studies and different systems, in multilayers by example in \cite{McIntyre97} or in Co thin films deposited on Pt(111) single crystal \cite{SaintLager1998}. Our results also shows that the annealing not only promotes interdiffusion but also locally the chemical ordering. However considering the intensity of the 111 peak, the chemical order in our film is very low.

The existence of this alloying of Co and Pt during annealing and the existence of the \lunun{} phase has several consequences. First of all, it is necessary to recalculate the value of $K_u$ and $M_s$ because we evaluated them by taking into account only the thickness of Co deposited. When replacing the thickness of the Co by the total thickness of the film we obtained :
\begin{eqnarray*}
K_u&=2.5\ 10^5\ \mathrm{J.m^{-3}}\\
M_s(400 \degrees)&=440\ \mathrm{kA.m^{-1}}\\
M_s(600 \degrees)&=496\ \mathrm{kA.m^{-1}}
\end{eqnarray*}

We may compare these values with the one found in the literature for CoPt films displaying the \lunun{} chemical ordering. Concerning the anisotropy, Iwata and coworkers\cite{Iwata1997} grew  $\mathrm{Co_{50}Pt_{50}}$ samples with $K_u=1\ 10^6\ \mathrm{J.m^{-3}}$, Sato \textit{et al.} did an extensive study growing $\mathrm{Co_{1-x}Pt_{x}}$ onto MgO with and without a Pt buffer. They obtained  $1.6\ 10^6\ \mathrm{J.m^{-3}}$ for approximately the same Pt contents ($30/46\approx 65\%$)and epitaxial films and $1.0\ 10^6\ \mathrm{J.m^{-3}}$ for approximately the same Pt contents and polycrystalline films\cite{Sato2008}. Sun \textit{et al.} obtained $3.2\ 10^5\ \mathrm{J.m^{-3}}$\cite{Sun2009}. Concerning the magnetization at saturation, Sato \textit{et al.}\cite{Sato2008} obtained 750 kA/m and Sun\textit{et al.} 740 kA/m \cite{Sun2009}. \textit{The lower value of $K_u$ in our films is easily explainable by their low chemical order. The shape (width and intensity) of the peak associated with the \lunun{} phase prevents any reliable evaluation of the chemical order factor \textit{S}}. Its presence however traduces the existence of chemical ordering correlation that are responsible for the appearance of PMA. Compared to previously reported results, our films are much thinner. Hence, this  allows a finer control of the coercive field because of the stronger demagnetizing field. Compared to ref.\citenum{Sun2009} results, our films show 100 \% remanence on a larger range of annealing temperature.

\begin{table}[h]
\begin{tabular}{|c|c|c|c|}
\hline & type&$K_u$ $(\mathrm{J.m^{-3}})$& $M_s$ (kA/m)\\
\hline Iwata\cite{Iwata1997}&epitaxial&$1\ 10^6$&\\
\hline Sato\cite{Sato2008} & epitaxial&$1.6\ 10^6$&750 \\
\hline Sato\cite{Sato2008} &polycrystal&$1.0\ 10^6$&750 \\
\hline Sun\cite{Sun2009} &polycrystal&$3.2\ 10^5$&740 \\
\hline This Study&polycrystal&$2.5\ 10^5$&440 \\ \hline
\end{tabular}
\caption{Summary of the magnetic properties of \lunun{} CoPt reported by different groups.\label{tab1}}
\end{table}

The \lunun{} phase is metastable and does not appear in the Co-Pt phase diagram. If the growth temperature is above 500\degrees{}, previous studies have shown that no \lunun{} phase is grown \cite{Iwata1997,Sun2009}. In our case we observe the disapperance of the \lunun{} phase upon annealing for \Ta above 550 \degrees. This is slightly higher than in ref.\citenum{Sun2009}.

\textit{Contrary to ref \citenum{Sun2009}, we observe the transformation from a \lunun{} phase to a \lunzero{} phase in our system around 550 \degrees. We have applied the Stoner-Wohlfarth model\cite{Aharoni} to our film to extract the characteristic parameter of the \lunzero{} phase. Indeed if we assume that we have independent grains with $<100>$ axis distributed on a cone making an angle of 54.7\degree{} with the film normal, we can extract the switching field of each grain. According to Stoner-Wohlfarth, that switching field is related to uniaxial anisotropy. From this relationship we derived a value for $K_u$. We obtain for \Ta=650\degrees{} a value of $K_u=8\ 10^5 \mathrm{J.m^{-3}}$. This is much lower than standard value for equiatomic epitaxial CoPt in the \lunzero{} phase. One explanation maybe that our films only contain 35 \% of Co atoms. This fact may also explain the lower value of $M_s$. }


%
%
%

Finally, we believe that the appearance of the \lunun{} phase upon annealing is an complementary explanation to the existence of PMA in the annealed trilayer \alumina/Co/Pt. It is more consistent with the strong interdiffusion of Co and Pt atoms upon annealing. This could provide an explanation to the unresolved observation of PMA in annealed MgO/CoFeB/Ta or MgO/CoFeB/Pd trilayer reported in the literature.


\section{Conclusion}

\alumina/Co(16\AA)/Pt(30\AA) films have been grown by sputtering with a (111) texture. After annealing at, at least, 350 \degrees, they exhibit PMA. These films are 100\% remanent. The coercive field of the layers increase linerarly with the \Ta. Above \Ta=550\degrees, the coercive field changes abruptly. XRD has demonstrated that the films display crystallites showing \lunun{} chemical ordering. As this phases is metastable, increasing the annealing temperature destroys the \lunun{} phase. However we see the appearance of grains having \lunzero{} chemical ordering. The [001] axis being isotropically oriented around the normal of the film, the existence of the \lunzero{} phase explains the strong increase of the coercive field above \Ta=600\degrees.
We believe that the existence of the \lunun{} phase in our films explains partially the appearance of the PMA upon annealing.

\newpage

\end{document}